\documentclass{elsart}
\journal{New Astronomy}
\usepackage{epsfig}
\usepackage{rotating}

\begin{document}

\begin{frontmatter}
\title{
Long-Term X-ray Variability in GX\,354--0
}

\author{A.~K.~H.~Kong\thanksref{ak}},
\author{P.~A.~Charles\thanksref{pc}} and
\author{E.~Kuulkers\thanksref{ek}}

\address{
Department of Astrophysics, Nuclear \& Astrophysics 
Laboratory, Keble Road, Oxford OX1 3RH, U.K.}

\thanks[ak]{E-mail: albertk@astro.ox.ac.uk}
\thanks[pc]{E-mail: pac@astro.ox.ac.uk}
\thanks[ek]{E-mail: erik@astro.ox.ac.uk}

\begin{abstract}
We report for the first time the detection of long-term X-ray variability in 
the bright bulge source  
GX\,354--0 (=4U\,1728--34) observed with the  All Sky Monitor (ASM) on board  
the {\it Rossi X-Ray Timing Explorer} ({\it RXTE}).  The 2-year {\it 
RXTE} ASM database reveals significant power at $\sim$\,72 days.  Similar 
behaviour was seen in the 6-year {\it Ariel 5} ASM database, but at a 
period of $\sim$\,63 
days.  The timescales and light curves resemble the $\sim$\,78 days modulation 
seen in Cyg X--2 and we therefore interpret this 
modulation in GX\,354--0 as a super-orbital effect.
\end{abstract}  

\end{frontmatter}


\section{Introduction}

The low-mass X-ray binary (LMXB) GX\,354--0 (=4U\,1728--34) is a well-known 
X-ray burster (e.g. Basinska et al. 1984) which has been 
classified as an atoll source based on {\it
EXOSAT} data (Hasinger \& van der Klis 1989).  Despite its detection in 
both soft  
and hard X-ray bands, its orbital period is still unknown.  Based on {\it 
Einstein} HRI and infrared observations, Grindlay \& Hertz (1981) 
claimed the association of GX\,354--0 with a heavily reddened globular 
cluster.  
However, the existence of this cluster was not confirmed by later infrared 
observations (van Paradijs \& Isaacman 1989).  Apart from the bursting 
activities, GX\,354--0 also exhibits complex behaviour on short time scales.  
Quasi-periodic oscillations (QPOs) at 363 Hz (2.75 ms) during the burst itself 
were recently discovered 
by the {\it Rossi X-ray Timing Explorer} ({\it RXTE}) (Strohmayer, Zhang \& 
Swank 1996; 
Strohmayer et al. 1996a; 1996b) providing the first evidence 
for a millisecond spin period in LMXB. A comprehensive discussion of these 
temporal and 
spectral characteristics can be found in Strohmayer et al. 
(1997).  GX\,354--0 is likely to be a typical LMXB and can be considered as a 
burster involving a rapidly rotating neutron star. 

However, with no
optical or infrared  counterpart having yet been determined, because of the 
high extinction in this direction as well as the presumably large 
distance, there are few constraints on the nature or evolutionary state 
of GX\,354--0. We therefore decided to exploit the long-term monitoring 
capabilities of the ASM on both the 
{\it Ariel 5} and {\it RXTE} satellites in order to study the X-ray behaviour of 
this source on timescales of weeks to months.  In this way we could 
search for long-term (super-orbital) periods similar to those seen in other bright 
LMXBs, such as Cyg X--2 and X1820--30 (e.g. Smale \& 
Lochner 1992), and which might be related to either the accretion 
geometry, disc or neutron star properties.

\section{Observations}

\subsection{{\it RXTE All Sky Monitor}} 

The All Sky Monitor (ASM) (Levine et al. 1996) on board the {\it RXTE} 
(Bradt, 
Rothschild \& Swank 1993) consists of three wide-angle shadow cameras 
equipped with proportional counters that provide regular and almost 
continuous intensity measurements of most bright X-ray sources 
in three different energy bands (1.5--3, 3--5 and 5--12 keV).  The time 
resolution allows the ASM to cover 80$\%$ of the sky 
every 90 minutes.  Combined with a 40$\%$ duty 
cycle  (Levine et al. 1996), any given source is usually scanned 5--10 times per 
day.  Both individual dwell and one-day average light curves 
are made available via the WWW by the {\it RXTE} ASM team.  Here, we use 
the dwell data in studying the detailed long-term light curve of GX\,354--0.

In this analysis, we used 6550 measurements of GX\,354--0 which were made 
by the {\it RXTE} ASM from 1996 February to 1998 February (see Fig. 1a).

\subsection{{\it Ariel 5}}

The {\it Ariel 5} ASM experiment (Holt 1976) consisted of a pair of X-ray 
pinhole cameras with position sensitive proportional counters (3--6 
keV) that covered 75\% of the sky during each orbit 
($\sim$\,100\,mins). From the archival 
HEASARC {\it Ariel 5} database, we obtained 2142 data points on 
GX\,354--0 spanning the period from October 1974 to March 1980 (see Fig. 2a).

\section{Analysis and Results}

The {\it RXTE} ASM light curve of GX\,354--0 (Fig. 1a) shows 
substantial flaring activity, with count rates peaking at $\sim$ 26 ASM 
counts s$^{-1}$.  In order to search for periodic phenomena in 
the data set, we used two different methods: the Lomb-Scargle periodogram 
(Lomb 1976; Scargle 1982) 
and phase dispersion minimization (PDM; Stellingwerf 1978).  
The Lomb-Scargle 
periodogram is a modification of the discrete fourier transform which is 
generalized to the case of uneven spacing.  However, it is most sensitive 
in those cases  where any 
modulation present is sinusoidal.  PDM is suitable even where the data is 
non-sinusoidal and works by dividing the data into phase bins 
and minimizing the dispersion within the bins.  The deepest minimum relates to 
the period of the strongest modulation.

By applying the Lomb-Scargle periodogram to the {\it RXTE} ASM data, a 
distinct peak is found
at 71.7 days.  We determined the uncertainties on this period by 
generating 10,000 artificial data sets with the same variance, amplitude 
and period as the true data.  The distribution in 
the resulting peaks leads to an error in the true period of $\sim$ 0.5 d 
(1$\sigma$).  In Fig. 1b ({\it upper panel}) we plotted the 99\% significance 
level.  This is determined by generating noise data sets with the 
same time 
intervals and variance as the true data and then performing the 
Lomb-Scargle periodogram on the resulting data sets.  The peak power in each 
periodogram (which must be purely 
due to noise) was then recorded.  This was also repeated 10,000 times for good 
statistics.  We conclude that our peak at 71.7 days is highly significant.  
This result is also independently confirmed by the PDM analysis (Fig. 1b, 
{\it lower panel}). We note, however, that the deepest minimum occurs at $\sim$ 
117 days, but this is due to observation gaps in the {\it RXTE} ASM light curve 
(Fig. 1a).  The folded light curve of our {\it RXTE} ASM data on the period 
of 71.7 days is shown in Fig. 1c; phase zero is defined by the first data point. 

The {\it Ariel 5} ASM data do not show the variation as clearly as  
the {\it RXTE} ASM 
data, but some modulation can be discerned.  We note that there are 
various data gaps which will affect the window function. 
Both the Lomb-Scargle periodogram and PDM (Fig. 2b, {\it upper and middle 
panels}) show a peak near 365 days, which is due to the yearly 
variation in the count rate from solar X-ray scattering or fluorescing from 
the Earth's 
atmosphere (Priedhorsky, Terrell \& Holt 1983).  The second strongest peak is 
at 63 $\pm$ 0.1 days (1-$\sigma$ level) and lies well above the 99\% confidence 
level.  The window function is also shown (Fig. 2b, {\it lower panel}) to indicate 
the alias 
periods that would be induced by the yearly variation and observation gaps.   
Fig. 2c shows the folded light curve of the {\it Ariel 5} ASM 
data on the period of 63 days, where phase zero is defined by the first 
data point.

We also analysed the data from the {\it Vela 5B} XC (Conner, Evans \& Belian 
1969) which monitored the sky from 1969 May to 1979 June in the 3--12 keV 
band. However, due to the low sensitivity of this instrument, no 
variability was detectable.

\section{Discussion}

We have detected a 71.7 $\pm$ 0.5\,d periodicity in the {\it RXTE} ASM 
light 
curve and a 63 $\pm$ 0.1\,d periodicity in the {\it Ariel 5} ASM light curve of 
the LMXB GX\,354--0.  Such a long period modulation has hitherto been 
rather uncommon in LMXB. From 
the long-term variability survey carried out by Smale \& Lochner (1992) using 
{\it Vela 5B}, only 3 out of 16 LMXBs (X1820--30, X1916--05, Cyg X--2) were 
found to 
have long-term periods.  In particular, X1820--30 and X1916--05 have periods 
of 176 days and 199 days, respectively, while recent 
observations by {\it RXTE} ASM confirm the long-term period of $\sim$\,78 days in 
Cyg X--2 
(Wijnands et al. 1996).  These long-term periods are designated {\it 
superorbital}, as the 
orbital periods are known for all three objects (11.3\,mins, 50\,mins and 9.8\,d, 
respectively).  In massive X-ray binaries, super-orbital periods are 
more common.  Cyg X--1, SS433, LMC\,X--4, Her X--1 and several other sources 
all 
have super-orbital periods in the range of 30--300 days (e.g. Priedhorsky \& 
Holt 1987).  The cause of these super-orbital periods is still a subject of 
debate, and possible explanations  
include the precession of a tilted accretion disk, neutron star precession, mass
transfer feedback and triple systems (see Priedhorsky \& Holt 1987 and
Schwarzenberg-Czerny 1992 for more details).  

In the case of LMXBs the  
{\it super-orbital} period is much more rare and there is a large spread 
in values of the ratio of super-orbital to 
orbital periods.  For X1820--30, X1916--05 and Cyg X--2, the ratio is 22100, 
5750 and 8 respectively, whereas in massive systems it is in the 
range of 10 to 100 (see e.g. Wijers \& Pringle 1998).  The long-term 
variation in LMXBs may instead be due to radiation driven warped accretion discs 
(e.g. Wijers \& Pringle 1998) or a 
disc instability in the system (Priedhorsky \& Holt 1987; Dubus et al. 1998).

Many orbital periods of bright galactic sources (including GX\,354--0) remain 
unknown mainly due to the heavy optical extinction and/or crowded regions.  While 
the known orbital periods of 
LMXBs range from 11\,mins to $\sim$\,10\,d, Cyg\,X--2 and 
X0921--63 
actually have long orbital periods ($\sim$\,10 days). We note that the 
{\it RXTE} ASM light curves of Cyg X--2 (Fig. 3a) and GX\,354--0 are 
rather similar.  This similarity is enhanced by our analysis of the much more 
extensive {\it RXTE} ASM database of Cyg X--2 that is available now. 
The Lomb-Scargle periodogram shows a 69 $\pm$ 0.4\,d period (Fig. 3b) in 
addition to the already noted $\sim$\,78\,d period (Wijnands et al. 1996).  
Fig. 3c 
shows the folded light curve of Cyg X--2 with a period of 69 days, where 
phase zero is defined by the first data point. A recent study of GX\,1+4 by 
Chakrabarty \& Roche 
(1997) has suggested that the orbital period of this source may exceed 100 
days or even 260 days.  However, since any orbital modulation should be  
uniquely stable, we conclude that our periods of 63\,d and 72\,d found for 
GX\,354--0  are ``super-orbital'' 
and {\it not} orbital.  Further optical/infrared campaigns are needed to 
reveal its orbital period and to make further progress in this area.

\section{Acknowledgements}
We are grateful to Guillaume Dubus for useful comments on the discussion 
section. This paper utilizies quick-look results provided by the {\it 
ASM/RXTE} team and data obtained through the HEASARC Online Service of 
NASA/GSFC.

\newpage
\pagestyle{empty}
\begin{figure*}
{\rotatebox{-90}{\psfig{file=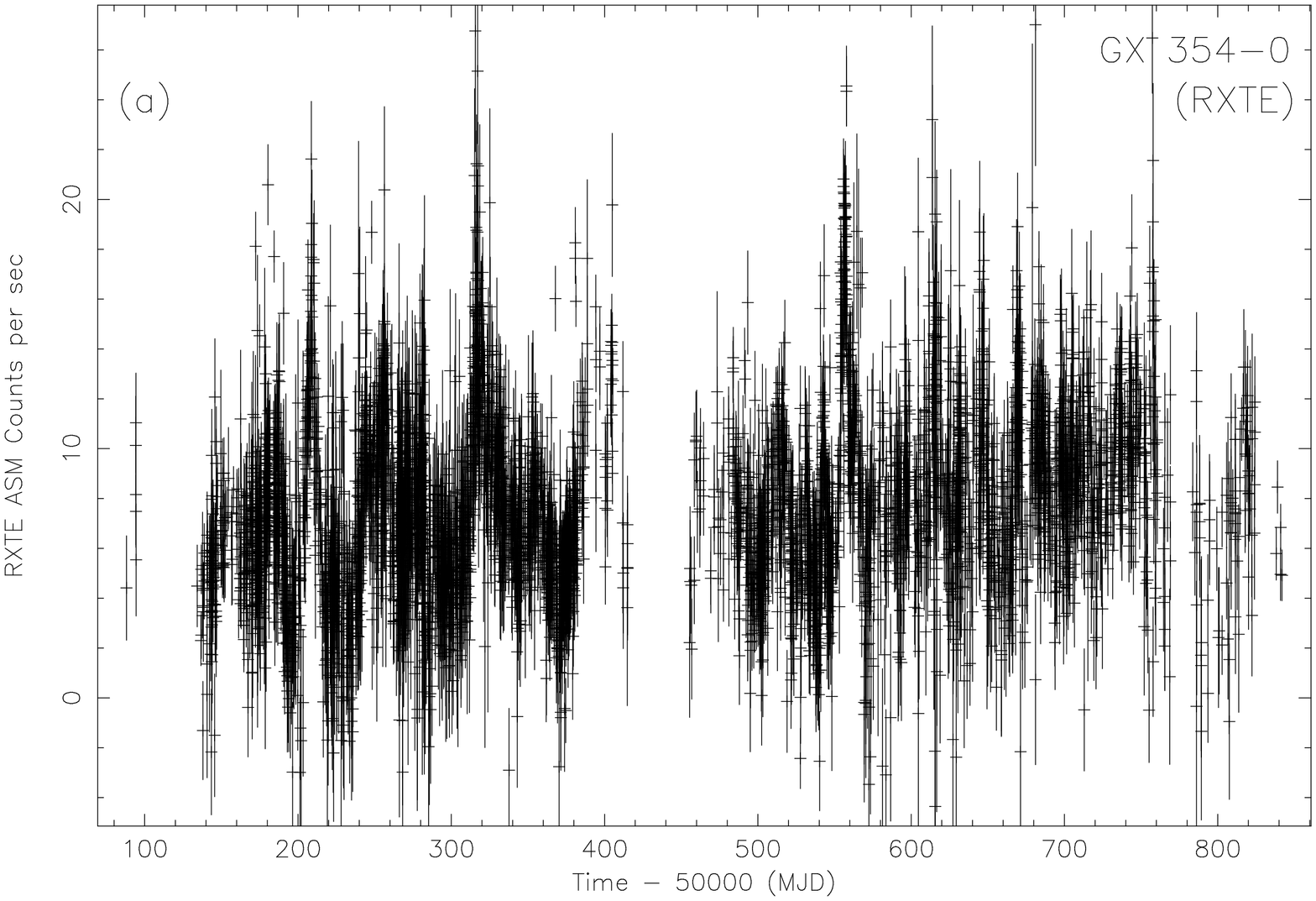,height=17cm,width=6.6cm}}}
{\rotatebox{-90}{\psfig{file=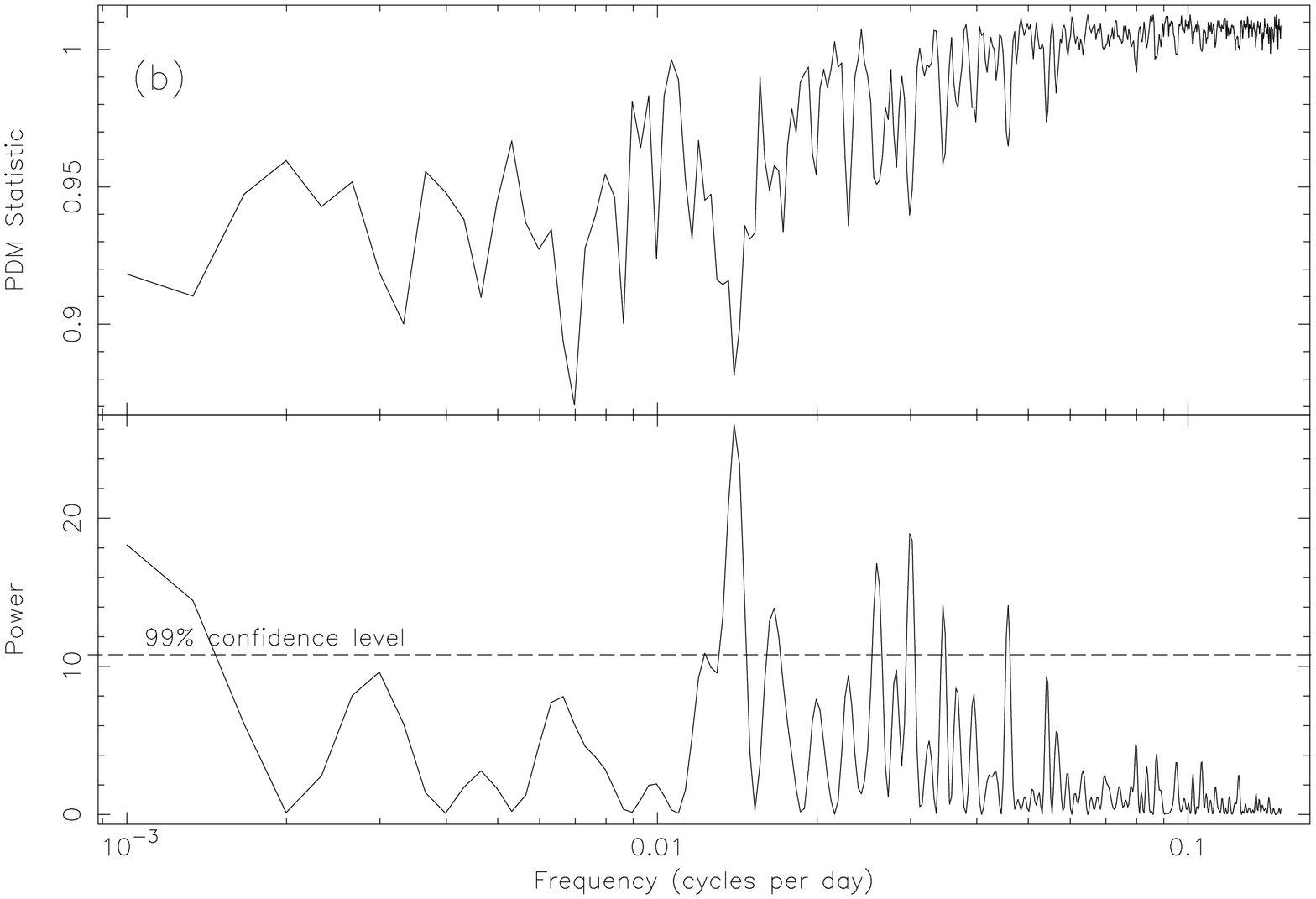,height=17cm,width=6.6cm}}}
{\rotatebox{-90}{\psfig{file=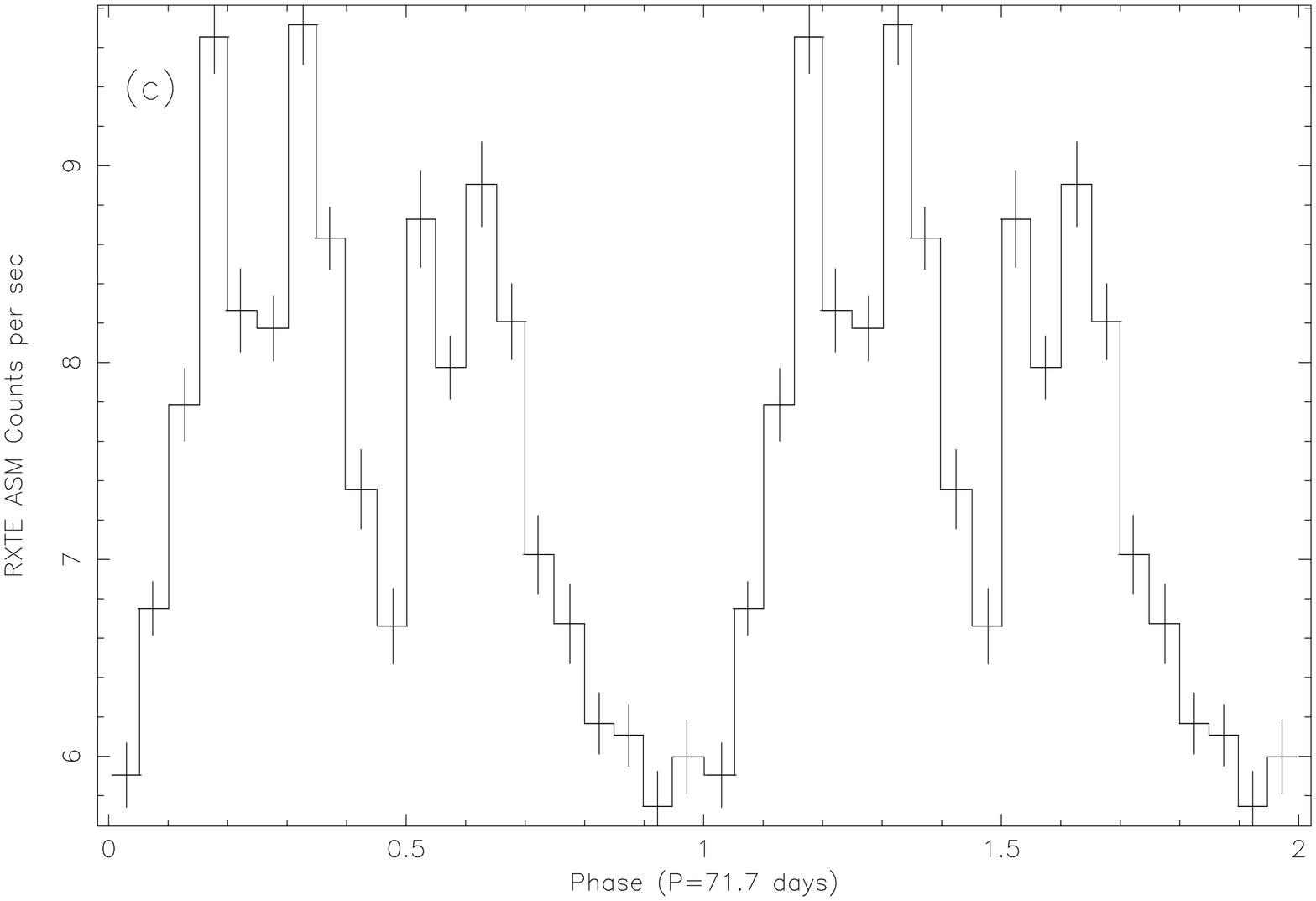,height=17cm,width=6.6cm}}}
\caption{\small{a) {\it RXTE} ASM 1.5--12 keV light curve of GX\,354--0
from 1996 February to 1998 February. b) PDM analysis ({\it upper panel}) of 
the {\it RXTE} ASM data, with the deepest minimum at $\sim$\,117 days and 
the next deepest at $\sim$\,72 days;
the Lomb-Scargle periodogram ({\it lower panel}) of the same data with a
peak at the same 72 day period.  The 99\% confidence level is shown by a dashed 
line. c) {\it RXTE} ASM folded light curve of GX\,354--0 on a
period of 71.7 days.  Two cycles are shown for clarity.  Phase zero is
arbitrary set at the time of the first data point (JD2450088.6)}} 
\end{figure*}

\begin{figure*}
{\rotatebox{-90}{\psfig{file=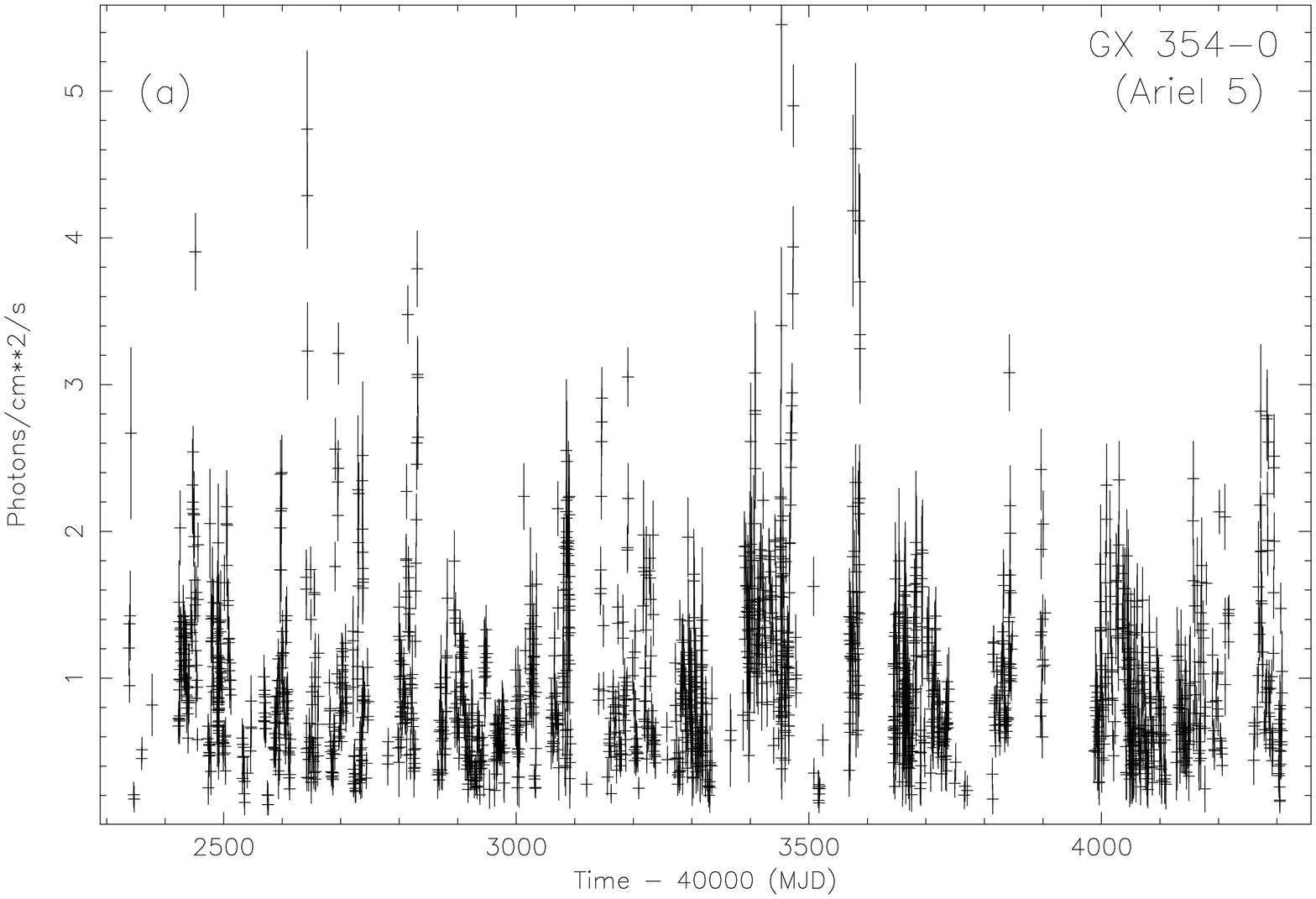,height=17cm,width=6.6cm}}}
{\rotatebox{-90}{\psfig{file=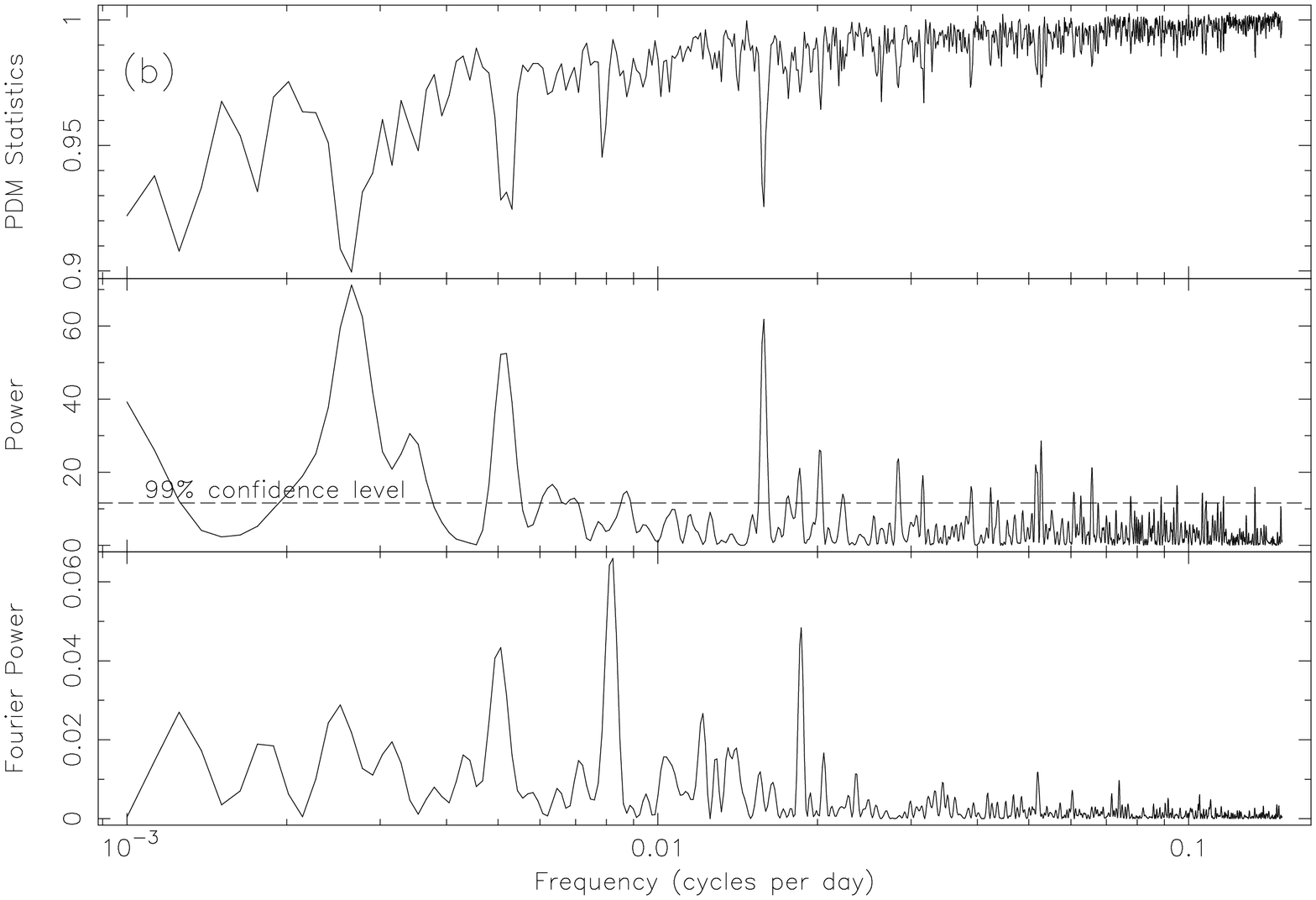,height=17cm,width=6.6cm}}}
{\rotatebox{-90}{\psfig{file=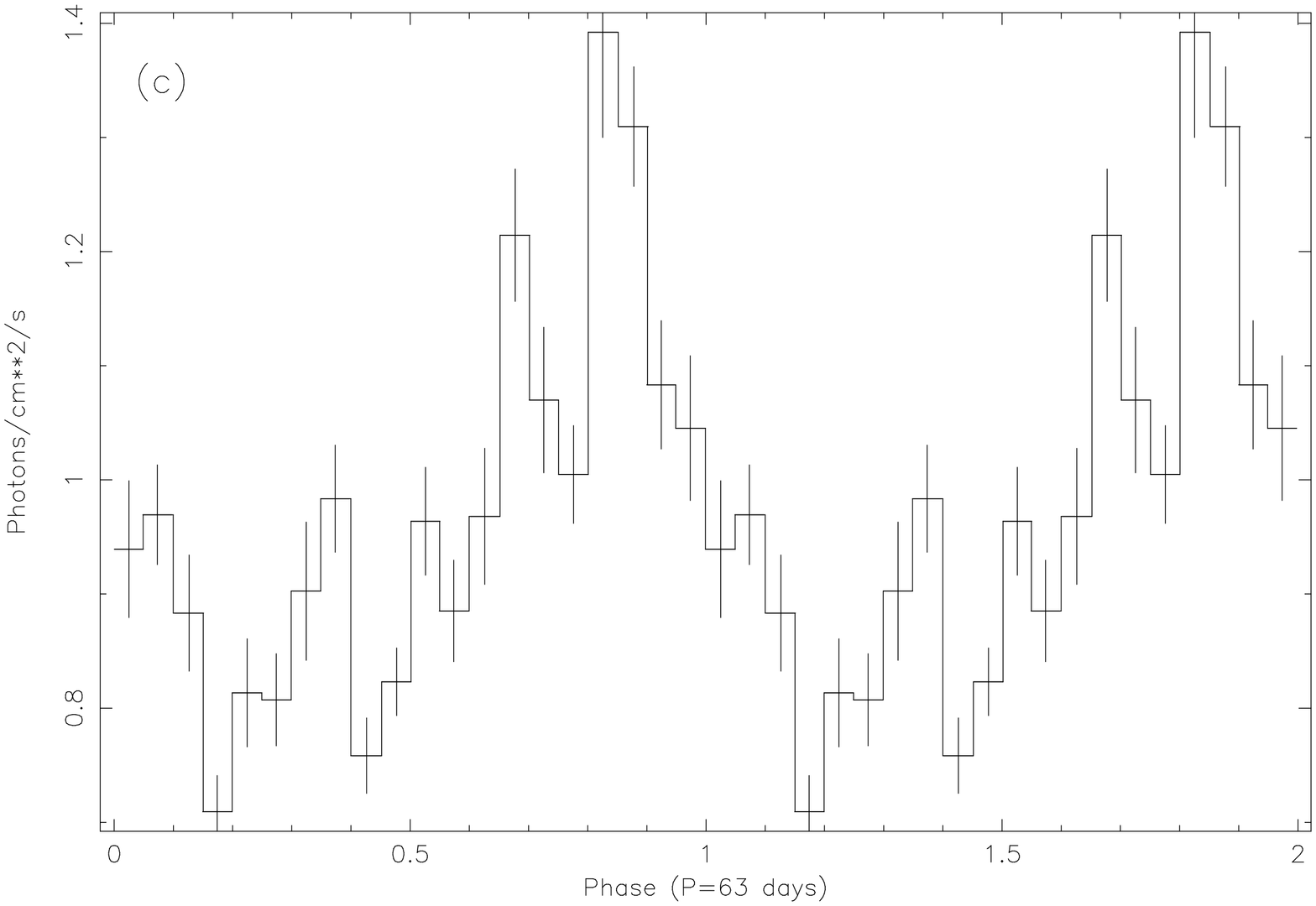,height=17cm,width=6.6cm}}}
\caption{\small{a) {\it Ariel 5} ASM 3--6 keV light curve of GX\,354--0 from 
1974 October to 1980 March. b) PDM analysis of the {\it Ariel 5} ASM data 
({\it upper panel}), with 
the deepest minimum at $\sim$\,365 days (due to solar scattering, see text) and 
the second deepest at $\sim$\,63 days; the 
Lomb-Scargle periodogram ({\it middle panel}) of the same data showing the 
same peak as the PDM at $\sim$\,63 days. The 99\% confidence level is shown 
as a dashed line.  
The window function of the {\it Ariel 5} ASM data ({\it lower panel}) 
clearly shows the alias peak in the two periodograms which are due to the 
observation gaps.  c) {\it Ariel 5} ASM folded light curve of GX\,354--0 on a 
period of 63 
days.  Two cycles are shown for clarity.  Phase zero is arbitrary set at 
the time of the first data point (JD2442338.6)}} 
\end{figure*}  

\begin{figure*}
{\rotatebox{-90}{\psfig{file=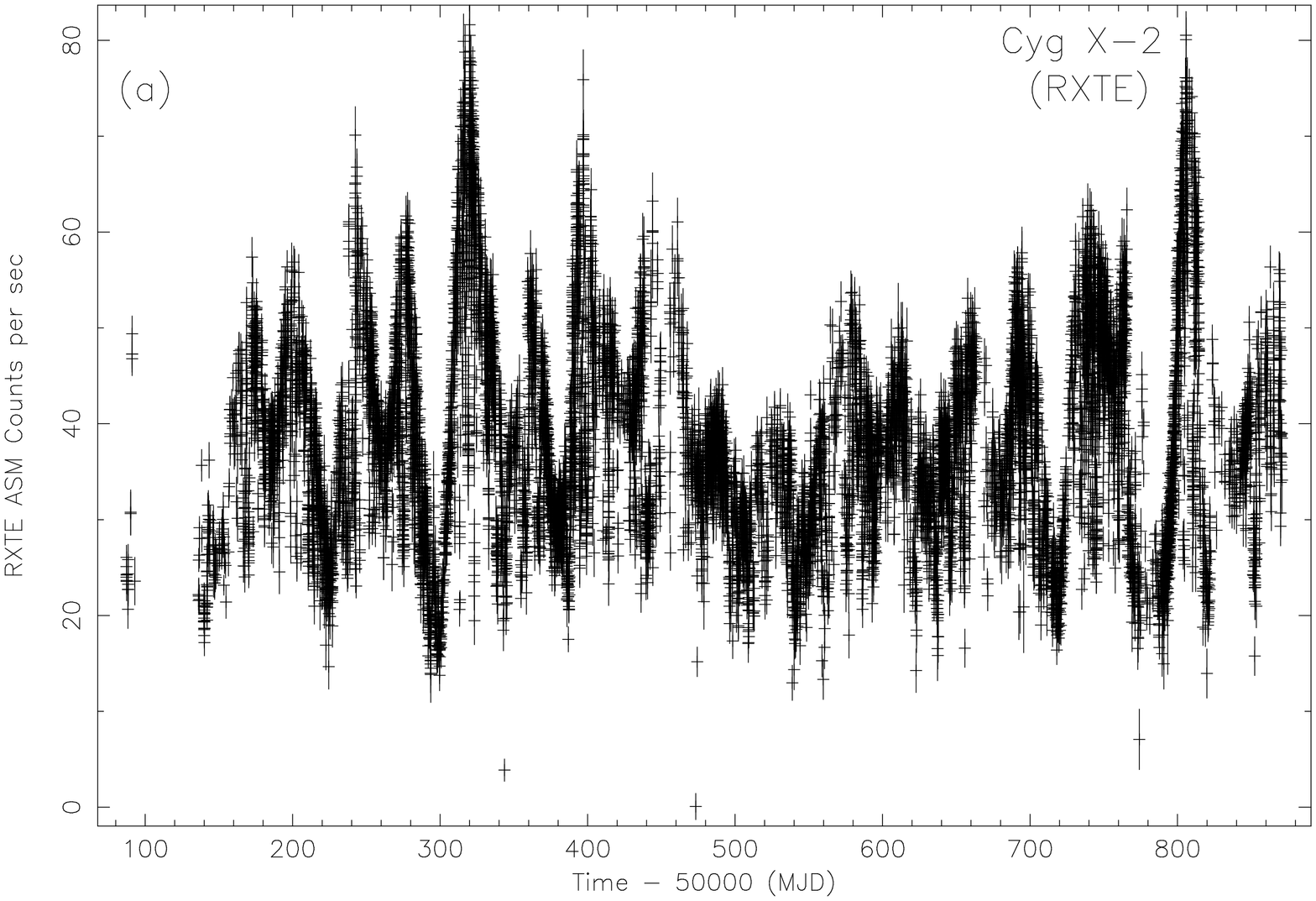,height=17cm,width=6.6cm}}}
{\rotatebox{-90}{\psfig{file=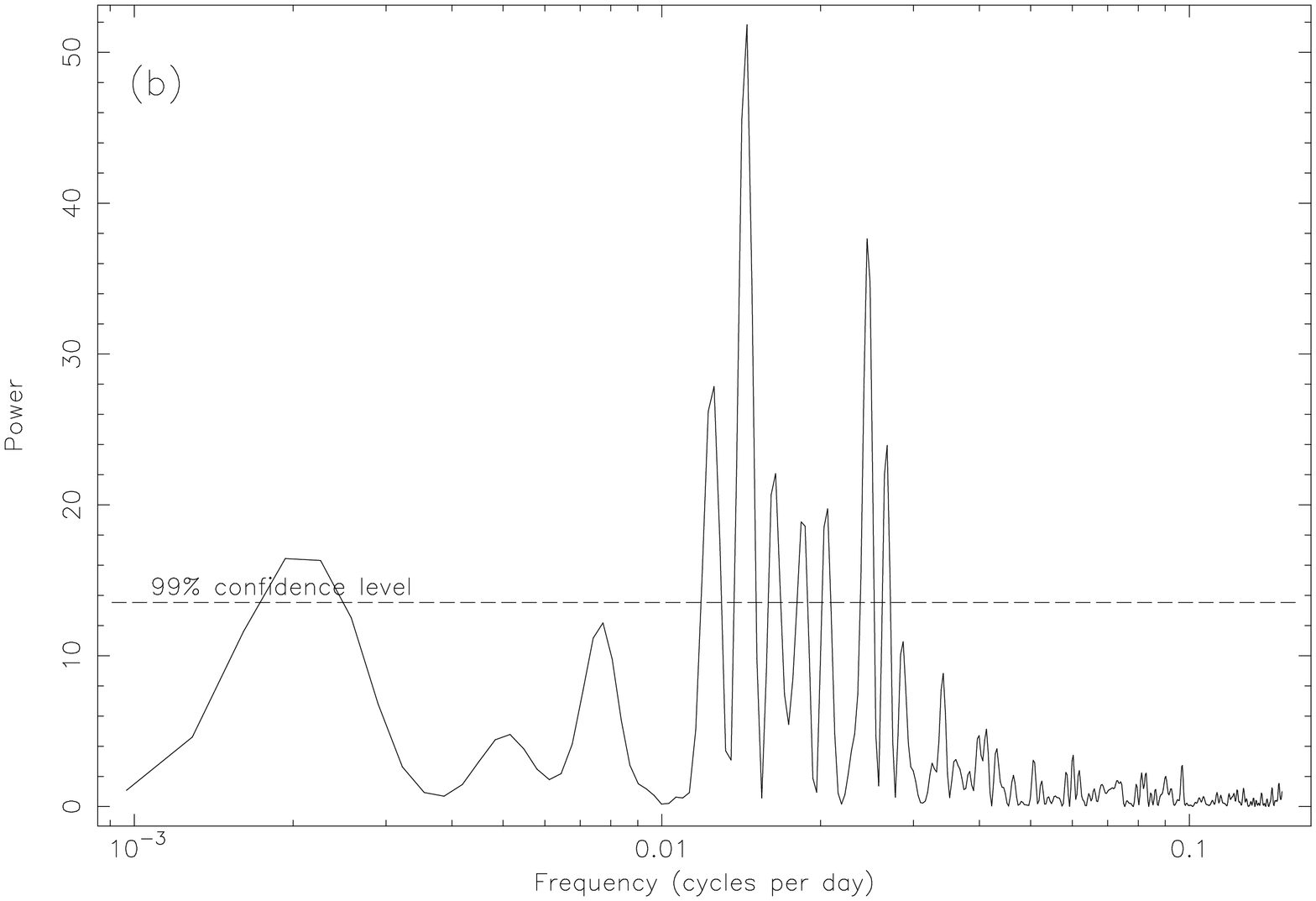,height=17cm,width=6.6cm}}}
{\rotatebox{-90}{\psfig{file=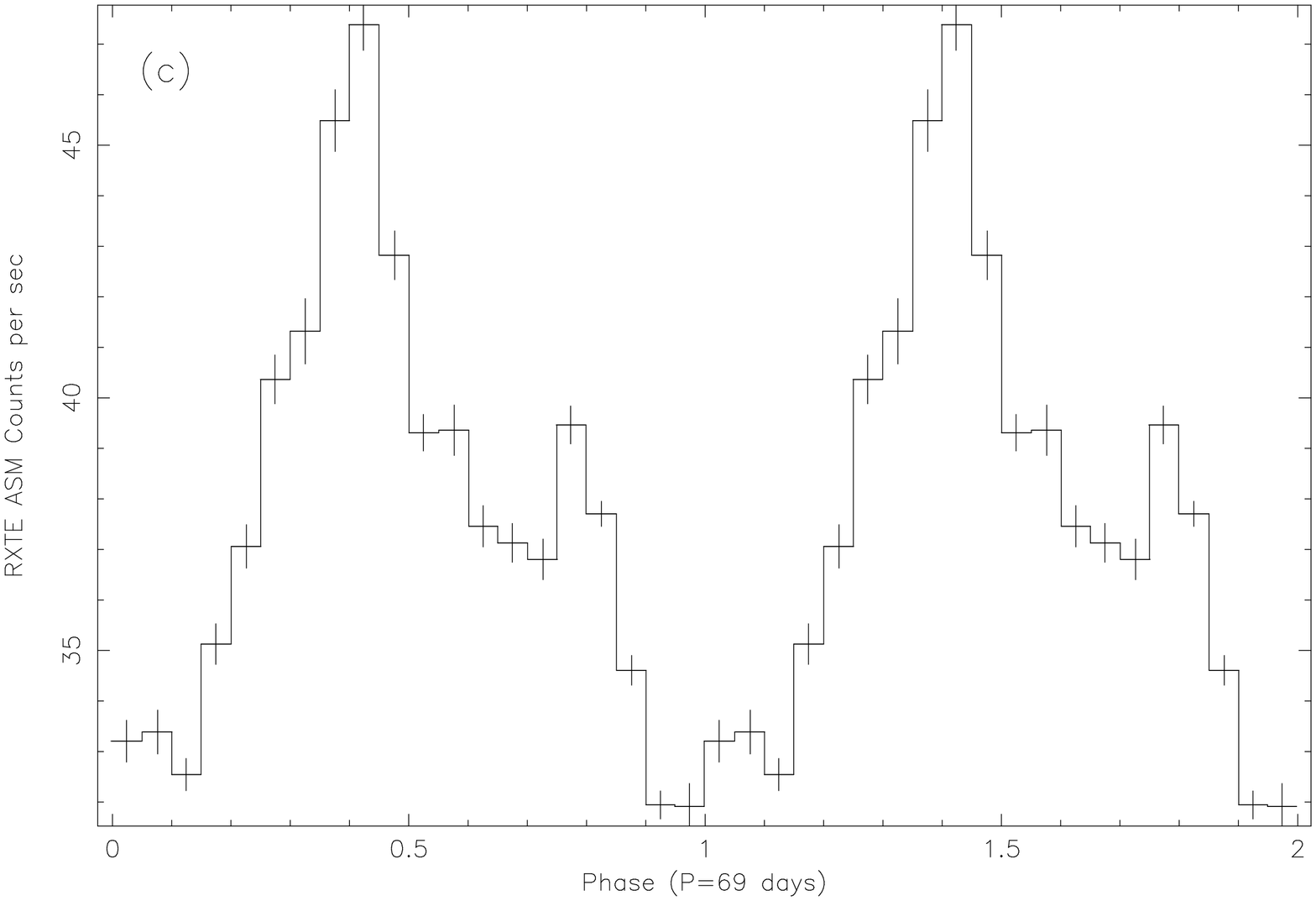,height=17cm,width=6.6cm}}}
\caption{\small{a) {\it RXTE} ASM light curve of Cyg X--2 from 1996 February 
to 1998 February.  b) Lomb-Scargle periodogram of the {\it RXTE} ASM data on 
Cyg X--2 showing a strong peak at $\sim$\,69 days. The 99\% confidence level 
is shown for reference.  c) Folded light curve of the {\it RXTE} ASM data 
of Cyg X--2 with period 69 days.  Phase zero is arbitrary set at the time of 
the first data point (JD2450087.8)}} 
\end{figure*}

\end{document}